\def\Title{From Boolean Valued Analysis to Quantum Set Theory:
Mathematical Worldview of Gaisi Takeuti\thanks{
This paper is an extended version of the paper published in 
S\={u}gaku Seminar 57 (2), 28--33 (Nippon Hyoron Sha, Tokyo, 2018) (in Japanese) by the
present author.
}}
\def\Author{Masanao Ozawa}
\def\AddressA{College of Engineering, Chubu University,
1200 Matsumoto-cho, Kasugai 487-8501, Japan}
\def\AddressB{Graduate School of Informatics, Nagoya University,
Chikusa-ku, Nagoya 464-8601, Japan}
\def\Abstract{Gaisi Takeuti introduced Boolean valued analysis around 1974 to provide systematic applications of Boolean valued models of set theory to analysis. Later, his methods were further developed by his followers, leading to solving several open problems in analysis and algebra. Using the methods of Boolean valued analysis, he further stepped forward to construct set theory based on quantum logic, as the first step to construct ``quantum mathematics'', a mathematics based on quantum logic. While it is known that the distributive law does not apply to quantum logic, and the equality axiom turns out not to hold in quantum set theory, he showed that the real numbers in quantum set theory are in one-to-one correspondence with the self-adjoint operators on a Hilbert space, or equivalently the physical quantities of the corresponding quantum system. As quantum logic is intrinsic and empirical, the results of the quantum set theory can be experimentally verified by quantum mechanics. In this paper, we analyze Takeuti's mathematical world view underlying his program from two perspectives: set theoretical foundations of modern mathematics and extending the notion of sets to multi-valued logic. We outlook the present status of his program, and envisage the further development of the program, by which we would be able to take a huge step forward toward unraveling the mysteries of quantum mechanics that have persisted for many years.}

\documentclass[12pt,oneside]{article}
  \let\Huge=\Large

  \let\Large=\large
  \let\large=\normalsize
\evensidemargin=\oddsidemargin
\usepackage[nospace]{cite}
\usepackage[dvipdfmx]{hyperref}
\usepackage{graphicx}
\usepackage{amssymb}
\usepackage{amsthm}
\usepackage{amsmath}
\usepackage{url}
\usepackage{doi}

\usepackage{color,bm}
\usepackage{enumerate}
\usepackage{bbm}
\usepackage{times}

\date{}
\title{\Huge\bf \Title}
\author{\sc\Author\\ \\
{\it\small \AddressA}\\
{\small\it \AddressB}}
\date{}
\author{\sc\Author\\ \\
{\it\small \AddressA}\\
{\small\it \AddressB}}
\date{}
  \newtheorem{Theorem}{Theorem}
  \renewcommand{\And}{\wedge}
  \newcommand{\B}{\cB}
  \newcommand{\R}{\rm{R}}
  \newcommand{\al}{\alpha}
  \newcommand{\be}{\beta}
  \newcommand{\om}{\omega}
  \newcommand{\ph}{\phi}
  
  \newcommand{\Iff}{\Leftrightarrow}
  \newcommand{\Inf}{\bigwedge}
 \newcommand{\Not}{\neg}
 \newcommand{\Or}{\vee}
  \newcommand{\Sup}{\bigvee}
  \newcommand{\VB}{\V^{(\cB)}}
\newcommand{\V}{V}
  \newcommand{\dom}{\mathop{\rm dom}}
  \newcommand{\bS}{\mathbf{S}}
  \newcommand{\cA}{{\mathcal A}}
  \newcommand{\cB}{{\mathcal B}}
  \newcommand{\cH}{{\mathcal H}}
  \newcommand{\cL}{{\mathcal L}}
  \newcommand{\cN}{{\mathcal N}}
  \newcommand{\cP}{{\mathcal P}}
  \newcommand{\cQ}{{\mathcal Q}}
  \newcommand{\cZ}{{\mathcal Z}}

\newcommand{\VQ}{\V^{(\cQ)}}
\newcommand{\val}[1]{[\![#1]\!]}
\newcommand{\cuniv}{\underline{\Or}}
\newcommand{\cdom}{\rotatebox[origin=c]{90}{$\models$}}
\newcommand{\On}{\rm On}
\newcommand{\Then}{\rightarrow}
\begin{document}
\maketitle
\begin{abstract}
\Abstract  \medskip

\newcommand\sep{,\ }
\end{abstract}

\section{Introduction}

In 1982, Gaisi Takeuti published a book entitled {\em Mathematical Worldview: Ideas and Prospects of 
Modern Mathematics} \cite{Ta82J}. In this book, he states that mathematics based on classical logic 
is an absolute truth, and it has the characteristic that all possible statements always belong to only one of
the two, either true or false,  and yet, in the future, the progress of human culture may lead to the birth of
a ``new mathematics'' based on a ``new logic.'' 
One well-known possibility is the creation of intuitionistic mathematics based on intuitionistic logic. 
This was advocated by Brouwer as one of the positions of mathematics, but Takeuti has 
proposed a ``new mathematics'' that should be called ``quantum mathematics'' based on quantum logic.
In this paper,  devoted to the memory of Takeuti, we will try explaining Takeuti's view of the mathematical 
world, the trajectory of research leading up to this proposal, and the current state of research based on this proposal.

\section{Quantum Logic}

 ``Quantum logic'' is the logic of quantum mechanics first discovered by Birkhoff
and von Neumann \cite{BvN36} in 1936. 
Among the laws of classical logic, it is known as the logic for which the distributive law does not hold.
Birkhoff and von Neumann hypothesized that the modular law holds true in place of the distributive law, 
but now a weaker orthomodular law is hypothesized.
On the other hand, in intuitionistic logic, the distributive law holds, but the law of excluded middle, the law of double negation, and De Morgan's law do not hold.
In this sense, quantum logic is a logic that contrasts with intuitionistic logic,
and Takeuti compared the relationship between classical logic, intuitionistic logic, and quantum logic to 
``God's logic,'' ``human logic,'' and ``the logic of things."

The fact that the distributive law does not hold in quantum logic is related to the existence of physical quantities that cannot be simultaneously measured in quantum mechanics due to the uncertainty principle.
Propositions considered in quantum logic are referred to as observational propositions. 
The most fundamental of these is the observational proposition that ``the value of the physical quantity $A$ 
is equal to the real number $a$,'' which is expressed as $A = a$. 
In quantum mechanics, a physical quantity is represented by a self-adjoint operator on a Hilbert space, 
and if it has no continuous spectrum, the possible values are its eigenvalues. 

For example, if the eigenvalues of such a physical quantity $A$ are $a_1$ and $a_2$ and the eigenvalues 
of a physical quantity $B$ are $ b_1$ and $b_2$, then $A=a_1\Or A=a_2$ and $B=b_1\Or B=b_2$ are 
true respectively.
However, if the physical quantities $A$ and $B$ have no common eigenvectors, for  any pair $(a_j, b_k)$ of eigenvalues of $A$ and $B$, the proposition $A = a_j\And $B$ = b_k$ is false. 
This means that 
\[
(A =a_1\Or A = a_2)\And (B = b_1\Or B = b_2)
\]
is true, but 
\[
(A =a_1\And B = b_1)\Or (A =a_1\And B = b_2)\Or (A = a_2\And B = b_1)\Or (A = a_2\And B = b_2)
\]
is false; therefore, the distributive law does not hold.

\section{Set Theoretical Worldview}

Constructing mathematics based on the logic characteristic of these kinds of
physical phenomena seems like an absurd undertaking at first. Where did this idea originally come from? 
According to Takeuti's {\em Mathematical World View} \cite{Ta82J}, there are two origins for this idea:
first, the idea that all the objects of mathematics are sets, and the modern mathematics equals ZFC set theory
 (an axiomatic set theory with Zermelo-Fraenkel's axioms plus the axiom of choice), and secondly, the idea
 that the concept of sets, originally considered in strictly two-valued logic, can be generalized 
 to multi-valued logic that allows intermediate truth values.
 
The first idea, which began with Frege, went through trials such as Russell's paradox, Hilbert's formalist program, 
and G\"{o}del's incompleteness theorem, but came to fruition in Bourbaki's  {\em \'{E}l\'{e}ments de
Math\'{e}matique} \cite{Bourbaki}.   It can be said that it is the fundamental idea of modern mathematics.

The second idea, which is famous for Zadeh's fuzzy set theory  \cite{Zad65,Ta92}, 
is to extend the two-valued logic of the
membership relation to a multi-valued logic in defining subsets of a given ``classical'' set. 
The method for expanding the universe of sets in the ordinary two-valued logic to multi-valued logic 
at once was clarified with Cohen's \cite{Coh63,Coh64} independence proof of the continuum hypothesis in the field 
of foundations of mathematics.

In this independence proof, Cohen \cite{Coh63,Coh64} developed a method, called forcing, of expanding 
the models of set theory.
Scott and Solovay \cite{SS67} showed that this method can be reformulated in a more accessible way 
using Boolean valued models of set theory.
The Boolean valued models of set theory make all the sets, 
as the objects of mathematics, to be "multivalued" 
into a multi-valued logic with truth values in a given Boolean algebra.
This view of multi-valued logicalization of the concept of 
sets further clarified the relationship between sheaf theory, topos theory, and intuitionistic set theory.

Therefore,  it can be said that Takeuti's idea of constructing mathematics based on quantum logic is to first develop this second idea, making all the sets that are the objects of mathematics to be multi-valued at once, 
and then under the "quantum set theory" obtained in the above,
to develop mathematics based on set theory along the first idea. 
Logic has two aspects, syntax and semantics.
Takeuti's program is to quantum-logicalize semantics at once, 
leaving the mathematical syntax that is based on the formal system of
set theory as it is.

\section{Modern Mathematics and ZFC Set Theory}

The majority of mathematicians accept that ZFC set theory is the foundation
on which modern mathematics is based. 
Takeuti wrote \cite[p.~171]{MT17}:
\begin{quote}
ZFC is a stable and powerful axiomatic system, there is almost no concern about contradictions, and all modern mathematics can be developed within it.
At present, it can be said that mathematics is equal to ZFC. 
\end{quote}
This is an important pillar of Takeuti's view of the mathematical world.

However, it is unfortunate that the 20th century,  in which these foundations of mathematics were established, became a little distant, and the interest in this idea and the achievements of Bourbaki, which was the basis for it, diminished. 
Let us add some comments.

Modern mathematics has two characteristics: formal and structural \cite{Rue07J}.
On the formal side, the whole mathematics can be developed using a formal language, strict deductive rules, and a set of axioms. 
 In other words, the whole mathematics can be formalized by a single first-order predicate theory, known as ZFC set theory,
and the only undefined term is the concept of sets.
The axioms are exhausted by the axioms of ZFC set theory, 
and all other mathematical concepts are defined within ZFC set theory. 
 For example, number systems such as natural numbers and real numbers, 
 geometric spaces such as Euclidean spaces,  
 and transfinite number systems of ordinal and cardinal numbers 
 are all defined inside ZFC \cite{Bourbaki}.

On the other hand, on the structural side, 
what mathematicians actually study is not limited to these kinds 
of numbers and figures, but instead it is generally considered as abstract objects called ``mathematical structures.'' 
Bourbaki showed that each field of mathematics is reduced to the study of "mathematical structures" specific to that field.  Those "mathematical structures" are defined within ZFC set theory,  and that their research is conducted based on the axioms of ZFC set theory \cite{Bourbaki}. 

While it is difficult to explain the concept of "mathematical structure" to the general readership, according to its formal definition, 
it is a set theoretical object that is given by a set 
with functions and relations defined on that set and 
several number systems (and the sets obtained 
by repeating direct products and power sets of these). 
Examples of mathematical structures include algebraic systems (such as groups, rings, and fields), topological space, 
manifolds, and measure spaces,
 as well as topological algebraic systems
(such as  Hilbert spaces and C*-algebras).

In this way, the word ``axiomatic system'' has two different aspects 
 corresponding to the formal and structural aspects of mathematics.
In other words, these are the axioms of  ZFC set theory,
and the axiomatic system that characterizes each mathematical 
structure and is included in its definition.
The concept of a mathematical structure gradually became clearer
 through the application of the axiomatic method proposed by Hilbert to various fields of mathematics, and resulted in Bourbaki's 
 {\em \'{E}l\'{e}ments de math\'{e}matique} 
\cite{Bourbaki}. 
 Each field of mathematics is relatively consistent from ZFC set theory if there is an example of the ``mathematical structure'' that the field studies.
 
Hilbert's formalism program can currently be interpreted as 
founding all mathematics in ZFC set theory and
having a finistic proof for the consistency of ZFC set theory.
Hilbert first showed that the consistency of geometry was reduced to the consistency of real number theory, and he questioned the proof of consistency of real number theory in the second problem of his
famous 23 problems, and he called for the axiomatizations of physics theories, including probability theory as the 6th problem.
As a result, probability theory was axiomatized by Kolmogorov \cite{Kol33}, by means of measure theory,
and became the basis for giving mathematical proofs to important hypotheses of statistical mechanics such as the ergodic hypothesis.
Measure theory is a study of the mathematical structure of measure space, and probability theory has been axiomatized as one of the ``mathematical structures'' 
in ZFC set theory.
In measure theory, the existence of Lebesgue non-measurable sets is a theorem of ZFC set theory, but it is known to be independent of set theory without the axiom of choice.
In addition, von Neumann's axiomatization of quantum mechanics shows that the two formulations of quantum mechanics, Heisenberg's matrix mechanics and Schr\"{o}dinger's wave mechanics, are unified by a common ``mathematical structure'' of the Hilbert space \cite{vN32}. 

Theory of operator algebras on Hilbert spaces 
plays an important role in the axiomatization of quantum field theory \cite{Kaw15}.
Recently, however, a problem independent of ZFC set theory have been identified within classical problems in operator algebras   \cite{Ake04}.
In these fields, we can get a glimpse of the reality of ``modern mathematics equals ZFC''.
However, Bourbaki's {\em \'{E}l\'{e}ments de math\'{e}matique} 
\cite{Bourbaki} does not include ``probability theory'' nor ``operator algebras''.
Nevertheless,  it is certain that they are the most Bourbaki-like fields in mathematics.

Returning to Hilbert's program, G\"{o}del's  incompleteness theorem showed that Hilbert's program is not feasible as it was.
So, is Takeuti's trust in ZFC, saying ``there is almost no concern about contradictions''
unfounded? 
 Putting aside the optimistic viewpoint that as long as contradictions are handled when they occur, 
 existing mathematics will not be lost, Bourbaki states the following regarding the proof of consistency \cite[p.~44]{Bou94}.
\begin{quotation}
The theorem of G\"{o}del does not however shut the door completely on attempts
to prove consistency, so long as one abandons (at least partially) the
restrictions of Hilbert concerning ``finite procedures''. It is thus that Gentzen
in 1936 \cite{Gen38}, succeeds in proving the consistency of formalised arithmetic,
by using ``intuitively'' transfinite induction up to the countable ordinal $\varepsilon_0$.
The value of the ``certainty'' that one can attach to such reasoning is without
doubt less convincing than for that which satisfies the initial requirements
of Hilbert, and is essentially a matter of personal psychology for each mathematician;
it remains no less true that similar ``proofs'' using ``intuitive''
transfinite induction up to a given ordinal, would be considered important
progress if they could be applied, for example, to the theory of real numbers
or to a substantial part of the theory of sets.
\end{quotation}
Actually, this achievement hoped for by Bourbaki came to pass afterward, and the results are already in our hands. 
Takeuti's fundamental conjecture \cite{Ta53}
and its solution \cite{Tak67,Pra68}
gave a Gentzen-style consistency proof for the theory of real numbers, and Arai \cite{Ara03,Ara04} gave a Gentzen-style consistency proof for a substantial part of ZFC set theory.

\section{Set Theory Based on Multi-Valued Logic}

According to Takeuti \cite{Ta81J}, 
the universe of set theory is composed of the following two principles.

{\bf C1.} Power set composition principle

{\bf C2.} Transfinite generation principle

Principle C1 states that for an arbitrary set, the totality of its subsets becomes
a set again. Principle C2 states that when a method for creating a new set is provided, the method can be repeated a transfinite number of times as much as possible. 
The most typical case of this principle is the generation of ordinal numbers, which starts from the empty set $\emptyset$ and endlessly repeats the operation of collecting the ordinal numbers constructed so far, and the whole ordinal number is generated, as follows.
$0=\emptyset,1=\{0\},2=\{0,1\},\ldots,\om=\{0,1,2,\ldots\},\om+1=
\{0,1,2,\ldots,\om\},\ldots$.
Sets like $\om=\{0,1,2,3,\ldots\}$ which do not contain the final ordinal number are called 
limit ordinal numbers. Then the order relation for two ordinals $\al,\be$ is defined as $\al<\be$ iff $\al\in\be$. 
Using these two principles, the universe $V$ of sets can be constructed as follows.

{\bf F1.} When ordinal number 0 arises, $V_0$ is composed as $\emptyset$.

{\bf F2.} When creating ordinal number $\al+1$ from ordinal number $\al$, $V_{\al+1}$ is composed as $\cP(V_\al)$.

{\bf F3.} When creating limit ordinal number $\al$, $V_{\al}$ is composed of $\bigcup_{\be<\al}V_{\be}$.

{\bf F4.} As long as the generation of ordinal numbers is continued, the composition of $V_{\al}$ will be continued endlessly. We call the union $\bigcup_{\al}V_{\al}$ of $V_{\al}$ over all the ordinal numbers $\al$,  as $V$.

To consider a general method for constructing a set theory based on a multi-valued logic,  consider the following general structure of semantics of a logic.  Let $\cL$ be a partially ordered set such that every subset $S$ has the supremum $\Sup S$ and the infimum $\Inf S$, and that for the arbitrary element $x\in\cL$, its complement $x^{\perp}$ is defined with the following properties: $x\And x^{\perp}=0$, $x\Or x^{\perp}=1$, $x^{\perp\perp}=x$, and $x\le y \Iff y^\perp\le
x^\perp$.  Here, we write $x\And y=\Inf \{x,y\}$, $x\Or y=\Sup \{x,y\}$.  The element $1=\Sup \cL$ is the maximum element 
of $\cL$ representing ``true'',  the element $0=\Inf \cL$ is the minimum element of $L$ representing ``false'', 
and the other elements of the set $\cL$ represent intermediate truth values. Such a structure $\cL$ is called a 
complete orthocomplemented lattice.

In the process of constructing the universe of sets, the concept of a power set appears in F2
as a sole procedure for creating a new set, so that to construct the universe of $\cL$-valued sets, 
it suffices to extend this part to the $\cL$-valued logic.
Incidentally, using the idea of fuzzy sets, for any set $X$, any function $A: D\to \cL$ from any subset $D$ of $X$ 
to $\cL$ can be considered as an $\cL$-valued subset of $X$.
Therefore, the universe $V^{(\cL)}$ of sets based on $\cL$-valued logic is defined as follows.

\begin{enumerate}
\itemsep=0pt
\item[(1)] $V_{0}^{(\cL)}=\emptyset$.
\item[(2)] $V_{\al+1}^{(\cL)}=\{u\mid u:\dom(u)\to \cL,\ \dom(u)\subseteq V_{\al}^{(\cL)}\}$.
\item[(3)] $V^{(\cL)}_\al=\bigcup_{\be<\al}V^{(\cL)}_{\be}$ if $\al$ is a limit ordinal.
\item[(4)] $V^{(\cL)}=\bigcup_{\al\in\On}V_{\al}^{(\cL)}$.
\end{enumerate}
Here, $\On$ stands for the set of ordinals. 

Define the $\cL$-valued truth values of any closed formulas
of the language of set theory augmented by the names of elements of $V^{(\cL)}$
as follows \cite{TZ73,Ta81,21QTD}.
\begin{enumerate}
\itemsep=0pt
\item[(1)] $\val{\Not \ph}=\val{\ph}^{\perp}$.
\item[(2)]  $\val{\phi\And \psi}=\val{\phi}\And \val{\psi}$.
\item[(3)]  $\val{(\forall    x\in u)\phi(x)}=\Inf_{x\in \dom(u)}(u(x)\Then
\val{\phi(x)}).$
\item[(4)]  $\val{(\forall    x)\phi(x)}=\Inf_{x\in V^{(\cL)}}\val{\phi(x)}$.
\end{enumerate}
Here, the operation $\Then$ on $\cL$  is defined by 
$P \Then Q=P^{\perp}\Or(P\And Q)$.
We consider the logical symbols $\Or$, $(\exists x\in y)$, and $(\exists x)$ to be  defined 
by 
\begin{enumerate}\itemsep=0pt
\item[(5)] $\phi\Or \psi :=\Not(\Not\phi\And \Not\psi)$.
\item[(6)] $(\exists x\in u)\phi(x):=\Not((\forall    x\in u)\Not\phi(x))$.
\item[(7)] $(\exists x)\phi(x):=\Not((\forall x)\Not\phi(x))$.
\end{enumerate}
The $\cL$-valued truth values of the membership and the equality relations between two elements $u,v$ of
$V^{(\cL)}$ are recursively defined as follows.
\begin{enumerate}\itemsep=0pt
\item[(8)] $\val{u\in v}=\val{\exists x\in v (x=u)}$.
\item[(9)] $\val{u=v}=\val{\forall x\in u(x\in v)\And \forall y\in v(y\in u)}$.
\end{enumerate}
As a result of the above, the $\cL$-valued truth value $\val{\ph}\in \cL$ is defined 
for any closed formula $\phi$ in set theory with constant symbols naming elements 
of $V^{(\cL)}$.

Let $V$ be the universe of the standard sets based on the 2-valued logic.
Then there exists an $\cL$-valued set $\check{a}$ corresponding to each $a\in V$.
In fact, $\check{a}$ is determined as an element of $V^{(\cL)}$ such that
$\dom(\check{a})=\{\check{x}| x\in a\}$ and that $\check{a}(\check{x})=1$ if $x\in a$.
Then the relationship between the standard sets $a$ and $b$ is isomorphic 
to the relationship between $\cL$-valued sets $\check{a}$ and $\check{b}$.

When $\cL$ satisfies the distributive law, 
\[P\And (Q\Or R)=(P\And Q)\Or (P\And R),\quad P\Or (Q\And R)=(P\Or Q)\And(P\Or R),\]
$ \cL$ is called a complete Boolean algebra. 
If $\cB$ is a complete Boolean algebra, $\VB$ is called a Boolean value model of set theory.
In this case, the following theorem holds \cite{TZ73}.
\begin{Theorem}[Scott-Solovay]\label{th:1} 
If $\phi(x_1,\ldots,x_n)$ is provable in ZFC set theory, 
\[\val{\phi(u_1,\ldots,u_n)} = 1\]
holds for any $u_1,\ldots,n_n\in \VB$.
\end{Theorem}

Let us show what kind of $\cB$ can be used to prove the independence of the continuum hypothesis (CH) 
(i.e., CH cannot be proved in ZFC set theory); note that the consistency 
 of CH (i.e., the negation of CH cannot be proved in ZFC set theory) was proved by G\"{o}del \cite{God40}.
Let $I$ be an index set of the cardinality larger than $2^{\aleph_0}$.
Let $X=2^{\aleph_0\times I}$ be the generalized Cantor space, a product topological space
of the direct product of $\aleph_0\times I$ copies of $2=\{0,1\}$.
Let $\cB(X)$ be the Borel $\sigma$-field of subsets of $X$. Let $m$ be a product measure
on $\cB(X)$ defined as follows.
\[
m(\{p\in X\mid p(j_1)=a_1,\ldots,p(j_n)=a_n\})=\left(\frac{1}{2}\right)^n,
\]
where $a_j\in\{0,1\}$ $(j=1,\ldots,n)$.
The existence of this kind of measure is based on Kolmogorov's extension theorem. 
Let $\cN$ be the collection of measure zero subsets.
Then the quotient Boolean algebra $\cB=\cB(X)/\cN$ is a complete Boolean algebra,
called the measure algebra of $m$. 
For this $\cB$, the value of $\val{\rm CH}$ can be calculated, and the value
$\val{\rm CH} = 0$ is obtained \cite[p.~173]{TZ73}. 
If CH can be proved from ZFC, then from Theorem 1,  $\val{\rm CH} = 1$ for any $\cB$, so that
 the independence of the continuum hypothesis is proved in this way.

\section{Boolean Valued Analysis}

Let $\ph$ be a logical formula in ZFC set theory representing a mathematical theorem.
According to theorem \ref{th:1}, the proposition ``$\val{\phi}=1$'' is a new, different theorem.
Because the proposition $\val{\phi} = 1$ is constructed by recursive rules, 
we can analyse its meaning, so that we will be able to restate $\val{\phi} = 1$ 
using familiar concepts.
If  $\psi$ is a mathematical proposition written with our familiar concepts, 
and if $\val{\phi} = 1$ and $\psi$ can be proved equivalent, 
$\psi$ will also become a new mathematical theorem. 
In other words, if we prove that $\val{\phi} = 1$ and $\psi$ are equivalent, then
rather than proving $\psi$ directly, we can prove  $\psi$ by proving $\phi$  instead.
In this case, $\phi$ is a far simpler proposition than $\psi$, 
and proving the equivalence of $\val{\phi} = 1$ and $\psi$ is similar 
to language translation in many ways. 
As a result, this approach is known to often lead to very promising prospects.

To apply this method to analysis, in {\em Two Applications of Logic to Mathematics} \cite{Ta78}
 Takeuti studied the structure of the real numbers in $\VB$ for a complete Boolean algebra $\B$ 
 of projections on a Hilbert space $\cH$ and showed that the real numbers in $\VB$ are
in one-to-one correspondence with the self-adjoint operators on $\cH$ such that 
their spectral projections belong to $\B$.
Here, the real numbers in $\VB$ is defined as the set of elements $u$ of $\VB$ satisfying
$\val{\R(u)}=1$, where $\R(x)$ is a logical formula in ZFC meaning ``$x$ is a real number''.
This shows that from theorems $\phi$ related to real numbers,
the theorem $\val{\phi} = 1$ for these kinds of self-adjoint operators can be systematically obtained.
For example,  from the theorem stating ``Every upper bounded set of real numbers has its supremum'', 
we obtain the theorem stating ``Every upper bounded set of mutually commuting self-adjoint operators has its supremum.''

This is a very powerful method, which Takeuti referred to as Boolean valued analysis. 
In his paper {\em Von Neumann algebras and Boolean valued analysis} \cite{Ta83a}, 
Takeuti took this a step further, showing that the class of all Hilbert spaces in $\VB$ 
are in one-to-one correspondence with the class of all normal *-representations 
of the commutative von Neumann algebra $\cA$ generated by $\B$; see also \cite{83BH}.
In addition, he further showed that the class of all von Neumann factors
(von Neumann algebras with trivial centers)  in $\VB$ are in one-to-one correspondence with the class of
von Neumann algebras such that their centers are isomorphic to $\cA$. 

Based on this finding, theorems for general von Neumann algebras are systematically derived from 
theorems on von Neumann factors.  
In operator algebras, there has been a well-known method, called the reduction theory,
for deriving theorems on general von Neumann algebras from theorems on 
von Neumann factors using the direct integral decompositions 
of von Neumann algebras into von Neumann factors, but there are restrictions such as  
separability.
The Boolean valued analysis method is a powerful approach which does not have these restrictions.

To show that Boolean valued analysis is a truly powerful method, the author explored 
the von Neumann factors in $\VB$ for the completely general complete Boolean algebra 
$\B$,
and showed that the type I von Neumann factors in $\VB$ are in one-to-one correspondence
with the type I AW*-algebras whose central projections are isomorphic to $\B$ \cite{83BT,84CT}; see also \cite{83BH,85NC}.
There is a theorem that ``every type I von Neumann factor is isomorphic to
the algebra of all bounded operators on a Hilbert space, and the cardinal number
representing the dimension of that Hilbert space is a complete invariant.''
Considering $\val{\ph} = 1$ for $ \phi $ to be the above theorem, we obtain the theorem
stating ``the isomorphic invariants of type I AW*-algebras corresponds to 
the cardinal numbers in $\VB$''. 

By the way, when $\cA$ is a type I von Neumann algebra,  
$\B$ satisfies the countable chain condition locally, 
so that the cardinal numbers in $\VB$ can be represented by the step functions of 
the standard cardinal numbers; this fact is known as the absoluteness of the cardinality
in $\VB$ for complete Boolean algebras $\cB$ satisfying the countable chain condition \cite[p.~162]{TZ73}.
In this case, it was already known in the theory of operator algebras 
that ``the step functions of the standard cardinal numbers forms 
an isomorphic invariant of the type I von Neumann algebras''.
However, it had been an open problem whether the isomorphic invariants of the 
type I AW*-algebras can be represented by the step functions of the standard
cardinal numbers, since Kaplansky \cite{Kap52} made a negative conjecture in 1952.
This conjecture was eventually settled in 1983 by the method of Boolean valued analysis \cite{83BT}.

In fact, the case where the cardinals in $\VB$ cannot be represented by the step functions 
of the standard cardinal numbers is known by the forcing method as ``cardinal collapsing'' \cite[Ch.~5]{Bel05}:
for any two infinite cardinals $\al$ and $\be$ in $V$, 
we can construct a complete Boolean algebra $\B$ such that $\check{\al}$ and $\check{\be}$ have the same cardinality in $\VB$.
Therefore, when the central projections forms such a complete Boolean algebra,
it is derived that the isomorphic invariants of those type I AW*-algebras cannot be 
represented by the step functions of the standard cardinal numbers.
Thus, Kaplansky's conjecture is settled \cite{83BT,84CT}.
For example, for arbitrary infinite cardinal numbers  $\al $ and $ \be $, we can construct 
 a commutative AW*-algebra $ \cZ $ such that the two type I AW*-algebras, 
 the $ \al\times\al $ matrix algebra over $ \cZ $  and $ \be\times\be $ matrix algebra over $ \cZ $, 
 are isomorphic \cite{85NC}.
 
In this way, Takeuti's Boolean valued analysis makes it possible to systematically 
apply the forcing method to analysis, and it
plays a role of a bridge between the two fields by applying the results in
foundations of mathematics to analysis. 
For further developments of Boolean valued analysis in operator algebras,
we refer the reader to \cite{Ta81,85TN,86BT,90BB,94FN}.

\section{Quantum Set Theory}

If $ \cL $ is a lattice consisting of the projections on a Hilbert space $ \cH $, then $ \cL $ is called the
standard quantum logic. 
In general, $\cL$  is called orthomodular iff $ P \le Q $ implies
that there exists a Boolean subalgebra of $\cL$ including $P$ and $Q$.
The standard quantum logic is a complete orthomodular lattice, 
and a complete orthomodular lattice is considered to be a general model of quantum logic.

Takeuti introduced the universe $V^{(\cQ)}$ of sets based on the standard quantum
logic $\cQ$ on a Hilbert space $\cH$ 
in his seminal paper {\em Quantum Set Theory} \cite{Ta81} published in 
1981, and started his research of quantum set theory. 

A remarkable fact, pointed out by Takeuti, about quantum set theory
is that the reals defined in $V^{(\cQ)}$ corresponds bijectively to  
the self-adjoint operators on $\cH$, as a straight forward consequence
of Boolean valued analysis based on complete Boolean algebras of projections,
developed by Takeuti in {\em Two applications of logic to mathematics}
\cite{Ta78}.  We refer to the above correspondence between the self-adjoit
operators and reals in $\VQ$ as the Takeuti correspondence. 

What formulas hold in $\VQ$?
To answer this question, Takeuti introduced the commutator $ \cdom (S)$ 
for any subset $S$ of $\cQ $  in \cite{Ta81},  and used it to define 
the commutator $\cuniv (u_1, \ldots, u_n)$ of any elements $ u_1, \ldots, u_n $ in $V^{(\cQ)} $.  
In addition, he showed that, roughly speaking,  
if one rewrites an axiom of ZFC by replacing $\forall x\ph(x)$ by 
$\forall x(\cuniv(x)\Then\ph(x))$ and replacing 
$\exists x$ by $\exists x (\cuniv(x)\And\ph(x)) $, the modified axiom
holds in $V^{(\cQ)}$. 
Then how theorems of ZFC hold true in $V^{(\cQ)}$? 
Using Takeuti's technique, the present author obtained the following theorem \cite{07TPQ,17A2,21QTD}.
\begin{Theorem}[Quantum Transfer Principle]
If  a theorem $\phi(x_1, . . . , x_n)$ of ZFC contains only bounded quantifiers,
then the relation
\[
\val{\ph(u_1,\ldots,u_n)}\ge\val{\cuniv(u_1,\ldots,u_n)}
\]
holds for any $u_1,\ldots,u_n\in V^{(\cQ)}$. 
\end{Theorem}
This theorem implies that any commuting family of quantum
sets $u_1,\ldots,u_n$ satisfies ZFC theorems, since in this case
$\val{\cuniv(u_1,\ldots,u_n)}=1$.  The above theorem  more generally 
states that any family of quantum physical quantities $u_1,\ldots,u_n$ 
satisfies ZFC theorems at least the truth value 
$\val{\cuniv(u_1,\ldots,u_n)}$.
 
In quantum mechanics, to every quantum system $\bS$ there corresponds
a Hilbert space $\cH$, and the physical quantities of $\bS$  are represented by 
the self-adjoint operators of $\cH$. 
Accordingly, $Q$ represents the logic of quantum system $\bS$, and 
the real numbers in $\VQ$ represent the physical quantities of $\bS$.
Based on this, all the observational propositions on a quantum system 
$\bS$  can be translated into logical formulas in ZFC referring to real 
numbers in $\VQ$ using the Takeuti correspondence \cite{16A2}. 
Therefore, mathematics based on quantum logic is nothing but a
mathematics in which the reals are quantum physical quantities.

A method for calculating the probability of observational proposition $\phi$
in quantum mechanics has been known as the Born rule.
By translating an observational proposition $\ph$ into the corresponding 
ZFC logical formula $\hat{\ph}$ using the Takeuti correspondence between physical quantities $A$
and reals $\hat{A}$ in $\VQ$, 
the probability calculation can be expressed as 
$\Pr\{\phi\| \psi\} = \|\val{\hat{\phi}}\psi\|^2$ \cite{16A2}. 
However, because the equality relation $A = B$ for two physical quantities 
$A$ and $B$ is not included in the observational propositions under the known quantum rule, 
the probability of $A$ and $B$ having the same value was not defined 
in quantum mechanics until recently. However, for the corresponding real numbers $\hat{A}$ and 
$\hat{B}$ in $\VQ$,  the truth value $\val{\hat{A} =\hat{B}}$ is defined in $\VQ$, so that by 
$\Pr\{A = B\|\psi\} = \|\val{\hat{A} =\hat{B}}\psi\|^2$, the probability of two 
physical quantities $A$ and $B$ having the same value has been newly defined \cite{16A2}. 
This theory was used to introduce the most basic condition in quantum
measurement theory requiring that ``the measured quantity $A$
and the meter quantity $B$ in the measuring instrument match''.
It  has contributed a great deal to reforming the uncertainty principle 
and other progress in this field \cite{16A2}. 
In this way, quantum set theory is a mathematical theory that has the power to extend conventional 
quantum mechanics, allowing it to be applied to new phenomena.

We have finally reached the starting point for constructing mathematics based on quantum logic,
we were able to clarify the equality relation between real numbers, 
and the study of the order relation between real numbers has just begun \cite{17A1,DEO20}. 
Since these relations show observable relationships that hold between 
the physical quantities of the quantum system, 
the results of quantum set theory can be immediately experimentally verified.
By inheriting and developing such a wonderful heritage of Takeuti's mathematical achievements, 
it must be possible to greatly advance the elucidation of the interpretational problems of quantum mechanics, 
which has been regarded as a mystery for many years.

\section*{Acknowledgments}
This research was supported by JSPS KAKENHI, No.~17K19970, and 
 the IRI-NU collaboration.

\end{document}